\documentclass[11pt]{article}
\usepackage[utf8]{inputenc}
\usepackage{amsfonts}
\usepackage{amssymb}
\usepackage{graphicx}
\usepackage{fancyhdr}
\usepackage{fancyvrb}
\usepackage{fancybox}
\usepackage{float}
\usepackage{geometry}
\usepackage{minitoc}
 \usepackage{indentfirst}
 \usepackage{multicol}
 \usepackage[outerbars]{changebar}
 \usepackage{pifont,textcomp}
 \usepackage{amsfonts}
 \usepackage{amsmath}
 \usepackage{amscd}
 \usepackage{array}
\usepackage{multirow}
 \usepackage{mathrsfs}
 \usepackage{amssymb}
 \usepackage{amsthm}
 \usepackage{color}
\usepackage{cite}
\usepackage{amsfonts}
\usepackage{amssymb}
\usepackage{graphicx}
\usepackage{fancyhdr}
\usepackage{fancyvrb}
\usepackage{fancybox}
\usepackage{float}
\usepackage{geometry}
\usepackage{minitoc}
\usepackage{indentfirst}
\usepackage{multicol, color}
\usepackage[outerbars]{changebar}
\usepackage{pifont,textcomp}
\usepackage{amsfonts}
\usepackage{amsmath}
\usepackage{amscd}
\usepackage{array}

\newtheorem{theorem}{Theorem}[section]
\newtheorem{lemma}[theorem]{Lemma}
\newtheorem{corollary}[theorem]{Corollary}
\newtheorem{proposition}[theorem]{Proposition}
\theoremstyle{definition}
\newtheorem{definition}[theorem]{Definition}

\newtheorem{Properties}[theorem]{Properties}
\theoremstyle{remark}
\newtheorem{remark}[theorem]{Remark}

\usepackage{multirow}
\usepackage{mathrsfs}
\usepackage{amssymb}
\usepackage{amsthm}
\usepackage{tikz}
\usetikzlibrary{matrix,arrows,decorations.pathmorphing}
\usetikzlibrary{shapes,arrows,positioning}

\advance \voffset by -0.8in
\advance \hoffset by -0.6in
\title{ Hilbert space representations  for Hermitian position-deformed Heisenberg algebra and Path integral formulation }
\author{ Thomas  Katsekpor$^{1,a}$, Latévi M. Lawson$^{2,3,4,b}$, Prince K. Osei$^{2,3,c}$ and Ibrahim Nonkan\'e$^{5,d}$  
	\space\\\\
	${}^1$Department of Mathematics,
	University of Ghana, \\	
 P.O. Box LG 62 Legon, Accra, Ghana\\\\
	${}^{2}$African Institute for Mathematical Sciences (AIMS) Ghana,\\
	 1 Shoppers Street, Manet, Spintex, Accra, Ghana\\\\
	 ${}^{3}$Quantum Leap Africa (QLA), AIMS Research Innovation Centre,\\ KN 3 Rd, Kigali, Rwanda.\\\\
	  ${}^4$Département de Physique, Faculté des Sciences, Université de Lomé,\\01 BP 1515 Lomé, Togo \\\\
	${}^5$Département d'\'Economie et de Math\'ematiques Appliqu\'ees,\\ IUFIC, Universit\'e Thomas Sankara, Burkina faso\\\\ 	 
tkatsekpor@ug.edu.gh$^a$,	latevi@aims.edu.gh$^b$, pkosei@aims.edu.gh$^c$, ibrahim.nonkane@uts.bf$^d$  }

\begin{document}
\maketitle

\begin{abstract}
 Position deformation of a Heisenberg algebra and Hilbert space representation of both maximal length and minimal momentum  uncertainties may lead to loss of Hermiticity of some operators that generate this  algebra. Consequently, the Hamiltonian operator constructed from these operators are also not  Hermitian. In the present paper, with an appropriate positive-definite Dyson map, we establish the Hermiticity of these operators by means of a similarity transformation. We then construct Hilbert space representations  associated with these  Hermitian operators that generate a  Hermitian Heisenberg algebra. With the help of these representations we establish the path integral formulation of any systems in this  Hermitian algebra. Finally, using the path integral of a free particle as an example, we demonstrate that the Euclidean propagator, action, and kinetic energy of this system are constrained by the standard classical mechanics limits.

\end{abstract}

\section{Introduction}
The study of Hilbert space representation of deformations for the uncertainty relation provides a promising approach to understand quantum gravity at the Planck scale \cite{1,2,3,4,5,6,7,8,9,10,11,12}.
They consist of quadratic Heisenberg algebra deformations in either momentum or position operators \cite{13,14,15,16,17,18,19,20,21}. It is well known that these deformations 
lead to maximal  and  minimal  uncertainties  and induce among other consequences  a loss of Hermiticity   of some operators that generate this algebra \cite{13}. Consequently,  Hamiltonians
$\hat H$ of systems involving these operators will in general also not be Hermitian. An
 immediate difficulty that arises when $\hat H$ is not Hermitian is that, the time evolution operator $ \hat U(t)=e^{-\frac{i}{\hbar}\hat H}$ is not unitary with respect to the inner product, resulting in non-conservation of the inner product under this time-evolution.

Non-Hermitian Hamiltonian  systems with real spectra in this context has been well studied in the  past few decades \cite{22,23,24,25,26,27,28,29,30,31,32,33,34,35,36,37,38,39,40,41}. 
 The quasi-Hermiticity \cite{24,31,32,39,40,41,42,43,44,45} and the pseudo-Hermiticity  \cite{29,33,34,35,40,44,45,46,47} are   synonymous used concepts that  allow  a consistent quantum mechanical description of these  systems.  They  are an unconventional approach to quantum mechanics,
  	based on the fact that  Hamiltonians (non-Hermitian
  	with respect to the conventional inner product of quantum mechanics) are  related to its adjoint via the existence of a metric operator. These  	Hamiltonians are in general assumed to be $\mathcal{PT}$-symmetric, that is, invariant under the joint action of space reflection ($\mathcal{P}$) and complex conjugation ($\mathcal{T}$), and they have often a real spectrum, usually discrete \cite{25,26}. 
       Despite their close relationship, both concepts are not always distinguished from one another.   
         	 In  quasi-Hermiticity, the metric operator is  linear, positive-definite,  Hermitian and not invertible \cite{24,31,32,42,43,44}  whereas
     	pseudo-Hermiticity refers to a linear Hermitian metric operator  which  is not necessary positive-definite  but  invertible \cite{33,34,35,40,44,46,47}.  However,  a given pseudo-Hermitian quantum system may or may not be quasi-Hermitian.  Then it is  quasi-Hermitian if the  space   includes a positive-definiteness of this metric  operator \cite{45,46}. While both methods enable consistent description of quantum mechanics, the fundamental problem with both approaches is the physical meaning of the Hilbert space that defines the metric operator and the Hamiltonian.  In order to map 	such theories in a consistent way to Hermitian theories, we present in this  paper
     	an alternative formulation that is based only on the unitary equivalence of  Hilbert spaces. This consists of defining 
     	an appropriate positive-definite Dyson map  \cite{23} which  establishes the Hermiticity of non-Hermitian Hamiltonian by means of a similarity transformation.  
      This provides a complete and an effective quantum mechanical descriptions with the conventional inner product.

A recently proposed quadratic  position-deformed Heisenberg algebra in 2D with simultaneous existence of
minimal and maximal length uncertainties  \cite{48}. It has been shown  that  this algebra  could be a
promising candidate  to probe quantum gravity \cite{48,49,51,52}. In the current work, we study the one dimensional case of this algebra which exhibits a maximal length and a minimal momentum uncertainties. As  has  been shown in \cite{48,49,51,52}, the deformation induces a loss of Hermiticity  of the momentum operator  which  consequently  forms with the position operator a non-Hermitian position-deformed Heisenberg algebra.
	 To raise up the  Hermiticity issue of this operator, 
	 we have proposed in \cite{52} an approach based on the introduction of a deformed completeness relation. This approach is similar to the pseudo-Hermicity quantum mechanics discussed in the latter paragraph. We construct the position wave function and its Fourier transform  that describes the corresponding Hilbert space.  However, in the present paper    with an appropriate positive definite Dyson map  we establish
  the Hermitian counterparts of the non-Hermitian operators by means of a similarity transformation.
   We  generate  with these Hermitian position and momentum operators, a  Hermitian position-deformed Heisenberg algebra that is isomorphic to the non-Hermitian one. The
position   wave function representation and its  Fourier transform representation associated with this Hermitian Heisenberg algebra are  constructed. By virtue of the additional correction term arising from the similarity transformation, we show that these Hilbert space representations provide an improvement  on the  one previously obtained in \cite{52}. We derive the propagators of path integrals and the classical action in these  representations.  It shows that, the action  which describe the classical trajectories of a  system defined by a  Hermitian Hamiltonian is  bounded by the ordinary ones of classical mechanics. It can be understood as follows: the classical system specified by the  Hermitian Hamiltonian  can travel in this space quickly because the quantum deformation effects shorten its paths.   The overall result achieved in this paper is no longer different  from the one obtained in \cite{52}.  Both results are correct and equivalent in the  sense that, the similarity transformation lets invariant the position-deformed Heisenberg algebra which is the generator of any  dynamical system in this space.
The equivalency of the position wave function representation and its Fourier transform demonstrate this, however the correction  factor
 separates both results. This correspondence is additionally noted in the formulation of the path integral. However, this path integral formulation generalizes the  standard   formulation in nonrelativistic quantum mechanics \cite{53} and provides an additional method to determine the propagator and the action of deformed quantum  systems. In fact,  the  standard formulation in the ordinary Heisenberg algebra suffers from a rigorous mathematical formulation when one takes into account  deformations of the paths. The current  formulation considers the situation where the classical paths are deformed 
 by  gravitational effects in quantum mechanics. When these effects are eliminated, the conventional path integral formulation is recovered. However, the path integral treatments of quadratic deformed quantum systems in the references\cite{54,55,56}  as well as \cite{57} are based on different approximation methods in the momentum representation, whereas in the present formulation, no approximation method is needed to treat the deformed systems.  Because it retains all of the information about the deformed parameter effects on the system dynamics, the current formulation is therefore more beneficial than the existing ones in the literature.

 This paper is outlined as follows: In section \ref{sec2}, we review fundamentals of  quasi- and  pseudo-Hermitians versus   Hermitian  quantum mechanics     and we comment on how the Hermitian  quantum mechanics provide an effective and conventional description of non-Hermitian quantum systems. 
 In section \ref{sec3},   we propose a metric operator $\tilde{S}_+$ and we show how the concept of pseudo-Hermicity is similar to the approach introduced in \cite{52} to solve the loss of Hermiticity due to the deformation.  We then deduce from this metric operator  a Dyson operator $G$ that establishes the   Hermitian  counterpart of pseudo-Hermitian operators in a Hilbert space with the standard inner-product.  By mean of similarity transformation, we generate  from the Hermitian operators,  a Hermitian  position-deformed  Heisenberg algebra. In section \ref{sec4}, we construct  Hilbert space representations  associated with this  ermitian  deformed  algebra. Section \ref{sec5} provides the path  integrals in these wave function representations and deduce the corresponding quantum propagators and classical actions. As an application, we compute the propagator, the action and the Kinetic energy of a Hamiltonian of a free particle and we show that these quantities are bounded by the ordinary ones without quantum deformation. 
In the last section, we present our conclusion.
\section{ Quasi- and Pseudo-Hermitians versus Hermitian quantum mechanics }\label{sec2}
\begin{definition}
 {\it   Let $\mathcal{H}$ be a finite dimensional Hilbert space  with the standard scalar product  $ \langle . |. \rangle. $ A  non-Hermitian Hamiltonian $\hat H:\mathcal{H}\rightarrow \mathcal{H}$  is
said to be  quasi-Hermitian \cite{24},} if there exists a  metric operator   $S:\mathcal{H}\rightarrow \mathcal{H}$  i.e.,  a positive-definite, Hermitian and  linear  operator such  that 
	\begin{eqnarray}\label{aaa12}
	 \hat H^\dag S = S\hat H.
\end{eqnarray}
\end{definition}
Since $ S$ is defined on the entire Hilbert space $\mathcal{H} $, 
$S$ is bounded.  This is known as the Hellinger-Toeplitz theorem\cite{58}.
As a consequence of the condition (\ref{aaa12}), the operator $\hat H$ eigenstates  no longer form an orthonormal basis and  the Hilbert space $\mathcal{H}$ structure needs to be modified.

\begin{definition} {\it
 A  Hilbert space $\mathcal{H}^{S}$ endowed  with a new  inner product $ \langle . |. \rangle_{S}$ in terms of the standard inner product $ \langle . |. \rangle $ is defined by
	\begin{eqnarray}\label{S111}
		\langle \psi|\phi \rangle_S:= \langle \psi| S\phi \rangle= \langle S^\dag\psi|\phi \rangle=\langle S\psi|\phi \rangle, \quad  \quad \psi,\phi\in \mathcal{H}.
	\end{eqnarray} }
\end{definition}
	For brevity we shall call the latter a quasi-inner
	product. Since the operator $S$ is positive-definite, one can easily show that $ \langle .|. \rangle_S$  is  positive-definite, non-degenerate and Hermitian \cite{59}. With the boundedness of $S$ one can show that $\mathcal{H}^{S}$   forms a  complete space \cite{46}  with the norm $ ||\phi ||_S=\sqrt{\langle \phi|\phi \rangle_S}$. In this way, the scalar product $ \langle .|. \rangle_S  $ can serve as the basis of a quantum theory.
 Note that this quadratic form \eqref{S111} reduces to the standard
Dirac inner product   when $S = \mathbb{I}$ as we would like, since in that case the system is described by
a Hermitian Hamiltonian.  Therefore, a Hermitian Hamiltonian defines  subset of  a quasi-Hermitian Hamiltonian.  A notion closely related to quasi-Hermiticity is pseudo-Hermiticity. In the current discussion of non-Hermitian Hamiltonians with real spectra, Mostafazadeh's work has highlighted its significance  \cite{28,44,46,47}.
	
\begin{definition}{\it   Let $\mathcal{H}$ be a finite dimensional Hilbert space  with the standard scalar product  $ \langle . |. \rangle.$ A  non-Hermitian Hamiltonian $\hat H:\mathcal{H}\rightarrow \mathcal{H}$  is
	said to be  pseudo-Hermitian \cite{46}, if there exists an automorphism $\tilde{S}:\mathcal{H}\rightarrow \mathcal{H}$ i.e.  an invertible, Hermitian, linear operator  satisfying
	\begin{eqnarray}\label{aaa1223}
		 \hat H^\dag = \tilde{S}\hat H\tilde{S}^{-1}.
	\end{eqnarray} } 
\end{definition}
Being an automorphism, its domain of definition
is the entire space, so that (again by virtue of the theorem of Hellinger and Toeplitz) it is bounded.  As a result, a quasi-Hermiticity and a pseudo-Hermiticity are the same concept, with the exception that $S$ must be positive-definite  not necessary invertible, in contrast to $\tilde{S}$ \cite{24}. To build a Hilbert space based on the scalar product \eqref{S111}, one must satisfy the criterion that $\tilde{S}$ must be positive-definite since it guarantees the positive-definiteness of the scalar product \eqref{S111}. Furthermore, according to  Def 2.3, the pseudo-Hermiticity of an operator is not sensitive to the particular form $ \tilde{S}$ of the operators  satisfying $ \hat H^\dag = \tilde{S}\hat H\tilde{S}^{-1}$  but to the existence of such operators. However, for a fixed operator $\tilde{S}$, the linear operator  $\hat H:\mathcal{H}\rightarrow \mathcal{H}$ satisfying  \eqref{aaa1223} is called $\tilde{S}$-pseudo-Hermitian \cite{28,44,45,46,47}.  As clearly explained in \cite{46}, $\tilde{S}$-pseudo-Hermitian operators are pseudo-Hermitian, but not every pseudo-Hermitian operator is
$\tilde{S}$-pseudo-Hermitian. This is because, $\tilde{S}$ may not be defined on the entire space $\mathcal{H}$. Finally, a pseudo-Hermitian Hamiltonian $\hat H$  may or may not be quasi-Hermitian \cite{45,46}. Then it is  quasi-Hermitian if the space $\mathcal{H}$  includes a positive operator $\tilde{S}_+$. Similarly,
the set of Hermitian operators is a proper subset of the set of quasi-Hermitian operators,
 quasi-Hermitian operators form a proper subset of the set of pseudo-Hermitian operators.  This can be schematically summarized as follows \cite{45,46} 
\begin{center}
	Hermitian $\subset$ Quasi-Hermitian $\subset$ Pseudo-Hermitian.
\end{center}
This provides a distinction between quasi-and pseudo-Hermiticity  which is not always made  \cite{60,61}.

  Moreover, one can ensure the conservation of the conventional probability interpretation of quantum mechanics with the use of this new inner product \eqref{S111}. To do this, we shall demonstrate that, relative to this inner product, the operator Hamiltonian is Hermitian.

\begin{proposition}
 {\it
 A non-Hermitian operator $\hat H $ is Hermitian with respect to the pseudo-inner product $ { \langle .|. \rangle_{\tilde{S}_+}}$ if we have
	\begin{eqnarray}
		\langle \psi|\hat H\phi \rangle_{\tilde{S}_+}:= \langle  \psi|\tilde{S}_+ \hat H \phi \rangle = \langle  \psi| \hat H^\dag {\tilde{S}_+}\phi \rangle ={ \langle \hat H \psi|\tilde{S}_+\phi \rangle=\langle \tilde{S}_+\hat H \psi|\phi \rangle=\langle \hat H \psi|\phi \rangle_{\tilde{S}_+}}.
	\end{eqnarray}}
\end{proposition}
	Operators, 
	such as  $\hat H$, which are Hermitian under the  pseudo-inner product $ \langle .|. \rangle_{\tilde{S}_+}$ are called {$\tilde{S}_+$-pseudo}-Hermitian operators \cite{45,46}.

\begin{lemma}
	 {\it
 Since the  Hamiltonian is Hermitian with respect to the inner product $ { \langle .|. \rangle_{\tilde{S}_+}}$,   this will result in conservation of probability under time evolution }
	\begin{eqnarray}
		\langle \psi(t)|\phi(t)\rangle_{\tilde{S}_+}&=&	\langle \psi(t)|S|\phi(t)\rangle=\langle \psi(0)| e^{\frac{i}{\hbar}t\hat H^\dag} {\tilde{S}_+}e^{-\frac{i}{\hbar}t\hat H} |\phi(0)\rangle \cr
		&=& \langle \psi(0)|{\tilde{S}_+}\left({\tilde{S}_+}^{-1} e^{\frac{i}{\hbar}t\hat H^\dag}{\tilde{S}_+}\right) e^{-\frac{i}{\hbar}t\hat H} |\phi(0)\rangle\cr
		&=& \langle \psi(0)|{\tilde{S}_+} e^{\frac{i}{\hbar}t\hat H} e^{-\frac{i}{\hbar}t\hat H} |\phi(0)\rangle\cr
		&=& 	\langle \psi(0)|\phi(0)\rangle_{\tilde{S}_+}.
	\end{eqnarray}
\end{lemma}
As we observe, the pseudo-Hermicity ensures  that the time evolution operator  $e^{-\frac{i}{\hbar}t \hat H}$  is unitary   with respect to this  inner product. However, the main issue with pseudo-Hermitian  quantum mechanics is related to the interpretation of physical space of the Hamiltonian observable \cite{28}. { An approach} 
 which improves the physical space  of this  observable  {  consists of defining its Hermitian Hamiltonian counterpart.  This can be achieved by 
 mapping the  pseudo-Hermitian  Hamiltonian defined in  $\mathcal{H}^{\tilde{S}_+}$ to its Hermitian counterpart} defined in  $\mathcal{H}$ { equipped with the standard  inner product $ \langle .|. \rangle$  by a similarity transformation. This transformation ensures   the unitary equivalence of both Hilbert spaces via   the existence a  }   Dyson map \cite{23}. 
{Given that $\tilde{S}_+$ is a positive-definite operator 	and  factorizing this  operator into
a product of a  Dyson operator $G$ and its Hermitian conjugate in the form  $\tilde{S}_+=G^\dag G$ allows  to define a Hermitian operator $\hat h $ counterpart to the pseudo-Hermitian operator $\hat H$.}

\begin{definition} {\it 
	An operator  $\hat h:\mathcal{H}\rightarrow \mathcal{H}$ is  Hermitian associated with the  pseudo-Hermitian operator $\hat H: \mathcal{H}^{{\tilde{S}_+}}\rightarrow \mathcal{H}^{{\tilde{S}_+}}$, if there exists  a  Dyson operator  operator $G:\mathcal{H}^{\tilde{S}_+}\rightarrow \mathcal{H}$, such that
	\begin{eqnarray}\label{ro}
		G \hat H G^{-1}=\hat h=\hat h^\dag.
			\end{eqnarray} }
\end{definition}

\begin{remark}

 i) It  follows from equation (\ref{aaa1223}) that
\begin{eqnarray}
	\hat H=G^{-1}(G^{-1})^\dag\hat H^\dag G^\dag G\iff G\hat H G^{-1}=(G^{-1})^\dag\hat H^\dag G^\dag=\left(G \hat H G^{-1}\right)^\dag,
\end{eqnarray}
where we can identify
\begin{eqnarray}
	G\hat H G^{-1}=	\hat h \quad \mbox{and}\quad \left(G \hat H G^{-1}\right)^\dag=\hat h^\dag
	\implies \hat h^\dag= \hat h.
\end{eqnarray}\\
ii) Schematically summarized, the latters can be described
by the following sequence of steps
\begin{eqnarray}
	\hat H&\neq&  \hat H^\dag \xrightarrow{{\tilde{S}_+}} {\tilde{S}_+} \hat {H} {\tilde{S}_+}^{-1}= \hat H^\dag \xrightarrow{G} G \hat {H}G^{-1}=\hat h=\hat h^\dag.
\end{eqnarray}	
\end{remark}
\begin{proposition}
	 {\it 	Let $\Phi,\Psi\in \mathcal{H}$  such that $\Phi=G\phi$ and $\Psi=G\psi$. There is  a unitary equivalency between 
 $\left(\mathcal{H}, \langle.|.\rangle \right)$ and 	$\left(\mathcal{H}^{\tilde{S}_+}, \langle.|.\rangle_{\tilde{S}_+} \right)$ \cite{47}. This is  shown  by
	\begin{eqnarray}\label{zq}
		\langle  \Psi| \Phi\rangle= \langle G  \psi|G \phi\rangle= \langle  \psi| G^\dag G \phi\rangle=\langle  \psi| {\tilde{S}_+} \phi\rangle=\langle  \psi| \phi\rangle_{\tilde{S}_+} \quad \mbox{with}\quad \phi,\psi \in \mathcal{H}^{\tilde{S}_+}.
\end{eqnarray} 
}
\end{proposition}
 Based on the  unitary equivalence  of the space  $\mathcal{H}^{\tilde{S}_+}$ and $\mathcal{H} $, we can show that the  operator  $\hat h$ is Hermitian  relative to the ordinary inner product  $ \langle . |. \rangle$. 
 
 \begin{lemma}
 	  {\it 	Let $\Phi,\Psi\in \mathcal{H}$.  An operator $\hat h $ is Hermitian with respect to  the inner product\, $\langle.|.\rangle $ if we have 
 \begin{eqnarray}
	\langle  \Psi| \hat h\Phi\rangle&=&\langle G^{-1} \Psi| G^{-1} \hat h\Phi\rangle_{\tilde{S}_+}=\langle G^{-1} \Psi| \hat H G^{-1} \Phi\rangle_{\tilde{S}_+}=\langle \hat H G^{-1} \Psi|  G^{-1} \Phi\rangle_{\tilde{S}_+}\cr
	&=& \langle G^{-1}\hat h   \Psi|  G^{-1} \Phi\rangle_{\tilde{S}_+}= \langle G^{-1}\hat h   \Psi|  G^{-1} \Phi\rangle_{\tilde{S}_+}=\langle \hat h  \Psi| \Phi\rangle.
\end{eqnarray}}
\end{lemma}
This is just a consequence of the unitary equivalence of the spaces $\mathcal{H}^{\tilde{S}_+}$ and $\mathcal{H}$. Consequently, its time-evolution operator $\hat u(t)=e^{-\frac{i}{\hbar}t \hat h}$ is unitary with respect to the ordinary inner product $ \langle .|.\rangle$ in $\mathcal{H}$.  

 \begin{corollary}
  {\it Let $\Phi,\Psi\in \mathcal{H}$ such that $\Phi=G\phi$ and $\Psi=G\psi$. Let $ u(t)=e^{-\frac{i}{\hbar}t \hat h}: \mathcal{H}\rightarrow \mathcal{H}$ be  time-evolution unitary operator,  we have
 \begin{eqnarray}
	\langle \hat u(t) \Psi|\hat u(t)\Phi\rangle&=& \langle \Psi|e^{\frac{i}{\hbar}t \hat h}e^{-\frac{i}{\hbar}t \hat h}|\Phi\rangle=  \langle \Psi|\Phi\rangle=
	\langle \psi|\phi\rangle_{\tilde{S}_+} \quad \mbox{with}\quad \phi,\psi \in \mathcal{H}^{\tilde{S}_+}.
\end{eqnarray}}
 \end{corollary}
\section{Hermitian position-deformed Heisenberg algebra }\label{sec3}

Let $\hat x_0=\hat x_0^\dag$ and $\hat p_0=\hat p_0^\dag$ be respectively Hermitian position and momentum operators defined as follows
\begin{eqnarray} \label{a1}
	\hat x_0\phi(x)=x\phi(x) \quad \mbox{and}\quad  \hat p_0 \phi(x)=-i\hbar\partial_x\phi(x).\label{3}
\end{eqnarray}
where $\phi(x)\in \mathcal{H}= \mathcal{L}^2\left(\mathbb{R}\right)$ is the one  infinite  dimensional (1D)    Hilbert  space.

 Hermitian  operators $\hat x_0$ and  $\hat p_0$  that act in $\mathcal{H}$  satisfy the  Heisenberg algebra
	\begin{eqnarray}\label{alg1}
		{[\hat x_0,\hat p_0 ]}=i\hbar\mathbb{I}\quad \mbox{and}\quad 	{[\hat x_0,\hat x_0 ]}=0=	{[\hat p_0,\hat p_0 ]}.
	\end{eqnarray}
	 The  Heisenberg uncertainty principle reads as
	\begin{eqnarray}
		\Delta x_0\Delta p_0\geq \frac{1}{2} \big| \large\langle\phi|[\hat x_0,\hat p_0]|\phi \large\rangle \big|\implies 
		\Delta x_0\Delta p_0\geq\frac{\hbar}{2}.
\end{eqnarray}

Let    $\mathcal{H}_\tau=\mathcal{L}^2\left(\Omega_\tau\right)$ be  a finite dimensional subset of  $\mathcal{H}$ such  that $\Omega_\tau\subset \mathbb{R}$ and $\tau\in (0,1)$ is a deformation
parameter. This parameter has been regarded in the references \cite{49,50,51,62} as the gravitational effects in quantum mechanics.  Let  $\hat X$ and $\hat P$ be respectively  position and  deformed momentum operators defined in $\mathcal{H}_\tau$ such that
	\begin{eqnarray}\label{a24}
		\hat X=\hat x_0\quad \mbox{and}\quad \hat P= \left(\mathbb{I}-\tau \hat x_0+\tau^2\hat x_0^2\right)\hat p_0.
	\end{eqnarray}
	 These  operators (\ref{a24})  form the following  position-deformed Heisenberg algebra \cite{37,50,51,62} 
	\begin{eqnarray}\label{alg4}
		[\hat X,\hat P]=i\hbar \left(\mathbb{I}-\tau \hat X+\tau^2\hat X^2\right), \quad 	{[\hat X,\hat X ]}=0=	{[\hat P,\hat P ]}.
\end{eqnarray}
From  the representation (\ref{a24}), it follows immediately that  the operator $\hat X$ is Hermitian while the operator $\hat P$ is  no longer Hermitian on the space $\mathcal{H}_\tau$ 
\begin{eqnarray}\label{ra}
	\hat X^\dag=\hat X\quad \mbox{and}\quad \hat P^\dag=\hat P-i\hbar\tau(\mathbb{I}-2\tau\hat X)\implies \hat P^\dag\neq \hat P, 
\end{eqnarray}
and when $\tau\rightarrow 0$, the momentum operator  $\hat P$  becomes Hermitian. The non-Hermiticity of the momentum operator $\hat P$ is induced by the deformation parameter $\tau$. This may be understood as if
 the quantum gravitational effects  are responsible  for the non-Hermiticity of this operator that generates the algebra \eqref{alg4}.  Furthermore, a Hamiltonian operator that includes this non-Hermitian operator  in  representation \eqref{a24}, is not a Hermitian operator as well and   nonconservation of the inner product under the time evolution
$\langle \psi(t)|\phi(t)\rangle\neq \langle \psi(0)|\phi(0)\rangle, \quad |\psi\rangle ,|\phi\rangle \in \mathcal{H}_\tau$.

In order to map 
 these operators (\ref{ra}) into the  pseudo-Hermitian ones,  we propose the metric operator $\tilde{S}_+$ given by 
	\begin{eqnarray}\label{b1}
	\tilde{S}_+= \left(\mathbb{I}-\tau \hat X+\tau^2\hat X^2\right)^{-1}.
	\end{eqnarray}
	It is easy to see that  the operator $\tilde{S}_+   $ is positive-definite  ($\tilde{S}_+ >0$), Hermitian ($\tilde{S}_+={\tilde{S}_+}^\dag$), and invertible. Since $\mathcal{H}_\tau$  is finite dimensional,  $ \tilde{S}_+ $ is bounded.     The pseudo-Hermicities are obtained by means of pseudo-similarity transformation 
	\begin{eqnarray}
		\tilde{S}_+ \hat X {\tilde{S}_+}^{-1}&=& \hat x_0= \hat X^\dag, \label{a21}\\
		\tilde{S}_+ \hat P  \tilde{S}_+^{-1}&=& \hat p_0\left(\mathbb{I}-\tau \hat x_0+\tau^2\hat x_0^2\right)= \hat P^\dag.\label{a22}
\end{eqnarray}
 Using equations (\ref{a21}) and (\ref{a22}), we obtain the pseudo-Hermicity of the Hamiltonian $\hat H$ such that
	\begin{eqnarray}\label{her}
		\tilde{S}_+ \hat H {\tilde{S}_+}^{-1}= \frac{1}{2m}\hat p_0\left(\mathbb{I}-\tau \hat x_0+\tau^2\hat x_0^2\right)\hat p_0\left(\mathbb{I}-\tau \hat x_0+\tau^2\hat x_0^2\right)+V(\hat x_0)= \hat H^\dag.
\end{eqnarray}
 A  Hilbert space $\mathcal{H}_\tau^{\tilde{S}_+}$ endowed  with a new  inner product $ \langle . |. \rangle_{\tilde{S}_+}$ in terms of the standard inner product $ \langle . |. \rangle $ is defined by
	\begin{eqnarray}\label{S1}
		\langle \psi|\phi \rangle_{\tilde{S}_+}= 
	\langle \psi|\tilde{S}_+\phi \rangle= \int_{\Omega_\tau}dx \psi^*(x)\left(\frac{\phi(x)}{1-\tau  x+\tau^2 x^2}\right)=\int_{\Omega_\tau}dx \left(\frac{\psi(x)}{1-\tau  x+\tau^2 x^2}\right)^*\phi(x)= \langle {\tilde{S}_+}^\dag \psi|\phi \rangle.
\end{eqnarray}
With the corresponding norm given by
	\begin{eqnarray}\label{a2}
		|| \phi||_{\tilde{S}_+}= \left(\int_{\Omega_\tau}\frac{dx}{1-\tau  x+\tau^2 x^2} |\phi(x)|^2\right)^{\frac{1}{2}}<\infty.
\end{eqnarray} 
{ We deduce from equations  (\ref{S1})  and (\ref{a2}), a    deformed completeness relation  introduced in \cite{lawson2023path}  to solve the non-Hermiticity of the momentum operator. It is given  by      
	\begin{eqnarray}\label{a221}
 \int_{\Omega_\tau}\frac{dx}{1-\tau  x+\tau^2 x^2} |x\rangle \langle x|=\mathbb{I}_{\mathcal{H}_\tau^{\tilde{S}_+}}.
\end{eqnarray} 	
With equation \eqref{a221} at hand, we can demonstrate that the  momentum operator $\hat P$ described in \eqref{a24} and the associated Hamiltonian  are Hermitian \cite{52} with regard to the pseudo-inner product $ \langle.|. \rangle_{\tilde{S}_+}$.
\begin{eqnarray}
	\langle \psi|\hat P\phi \rangle_{\tilde{S}_+}= \langle \hat P \psi|\phi \rangle_{\tilde{S}_+}\implies \langle \psi|\hat H\phi \rangle_{\tilde{S}_+}= \langle \hat H \psi|\phi \rangle_{\tilde{S}_+}.
\end{eqnarray} } 

Given that   ${\tilde{S}_+}=G^2$ is a positive-definite operator,  the positive-definite  Dyson map operator    is  simply
 	computed to be   
 \begin{eqnarray}\label{dyson}
		G=\sqrt{\tilde{S}_+}=\left(\mathbb{I}-\tau \hat X+\tau^2\hat X^2\right)^{-\frac{1}{2}}.
	\end{eqnarray}
	Thus, by means of  a similarity transformation of the above pseudo-Hermitian operators, the  Hermitian  counterparts $\hat x,\hat p$ and $ \hat h$ defined in $\mathcal{H}_\tau$ read as follows
	\begin{eqnarray}
		\hat x&=&G\hat X G^{-1}=\hat x_0=\hat x^\dag, \label{a9}\\ 
		\hat p&=&G \hat PG^{-1}=\left(\mathbb{I}-\tau \hat x_0+\tau^2\hat x_0^2\right)^{1/2}\hat p_0\left(\mathbb{I}-\tau \hat x_0+\tau^2\hat x_0^2\right)^{1/2}=\hat p^\dag, \label{a15}\\
		\hat h&=& G \hat HG^{-1} =
		\frac{1}{2m}\left(\mathbb{I}-\tau \hat x_0+\tau^2\hat x_0^2\right)^{\frac{1}{2}}\hat p_0\cr&& \times\left(\mathbb{I}-\tau \hat x_0+\tau^2\hat x_0^2\right)\hat p_0\left(\mathbb{I}-\tau \hat x_0+\tau^2\hat x_0^2\right)^{\frac{1}{2}}+V(\hat x_0)=\hat h^\dag. \label{a12}
	\end{eqnarray}
{The  Hilbert space $\mathcal{H}_\tau$ endowed  with the standard  inner product $ \langle . |. \rangle$  is defined by
	\begin{eqnarray}\label{zq}
		\langle  \Psi| \Phi\rangle= \int_{\Omega_\tau}dx \Psi^*(x)\Phi(x)   \quad \mbox{with}\quad \Phi,\Psi \in \mathcal{H}_\tau,
\end{eqnarray} 
 and the    completeness relation is given  by        
\begin{eqnarray}\label{a22}
	\int_{\Omega_\tau}dx |x\rangle \langle x|=\mathbb{I}_{{\mathcal{H}_\tau}}.
\end{eqnarray} 	
For any operator $\hat A,$ 
the expectation value and  the corresponding dispersions are given by
\begin{eqnarray}
	\langle \hat A  \rangle=	\langle \Phi|\hat A |\Phi \rangle\quad \mbox{and}\quad 
	\Delta A = \sqrt{\langle \hat A^2  \rangle-\langle \hat A \rangle^2}.
\end{eqnarray}

 }
 Hermitian operators (\ref{a9},\ref{a15})  generate a  Hermitian position-deformed Heisenberg algebra   similar to  the non-Hermitian one (\ref{alg4}) such that
	\begin{eqnarray}\label{alg5}
		[\hat x,\hat p]=i\hbar \left(\mathbb{I}-\tau \hat x+\tau^2\hat x^2\right), \quad 	{[\hat p,\hat p ]}=0=	{[\hat p,\hat p ]}.
	\end{eqnarray}

For a system of operators satisfying the commutation relation in (\ref{alg5}), the generalized uncertainty principle
is defined as follows
\begin{eqnarray}\label{leq11}
	\Delta  x\Delta p\geq \frac{\hbar}{2}\left(1-\tau\langle \hat x\rangle+\tau^2\langle \hat x^2\rangle\right), 
\end{eqnarray}
where $\langle \hat x\rangle$ and $\langle \hat x^2\rangle$ are the  expectation values of the operators $\hat x$ and $\hat x^2$ respectively for any  space representations.
Referring to \cite{37,48,49,50,62}, this equation leads to the absolute minimal uncertainty $\Delta p_{min}$ in $p$-direction  and the absolute maximal uncertainty  $\Delta x_{max}$ in $x$-direction  
when $\langle  \hat x\rangle=0$  such that
\begin{eqnarray}\label{q2}
	\quad \Delta x_{max}=\frac{1}{\tau}=\ell_{max}\quad  \mbox{and}\quad
	\Delta p_{min}=\hbar\tau.
\end{eqnarray}
This provides the  scale for the maximum length and minimum momentum obtained in \cite{49,51,62} which are
 different from the  condition imposed in  \cite{52}. As we shall see,  in contrast to the earlier conclusion in \cite{52}, the use of these uncertainty values in the current study has no impact on the physical interpretation.

\section{Hilbert space representations}\label{sec4}
Let $ \mathcal{H}_\tau=\mathcal{L}^2(\Omega_\tau)=\mathcal{L}^2(-\ell_{max},+\ell_{max})\subset \mathcal{H}$ be the Hilbert space representation in the spectral representation of these uncertainty measurements. We construct in this section the position space representation on  one hand  and the Fourier transform and its inverse representations on the other hand. 

\subsection{Position space representation}
\begin{definition}
	  {\it Let us consider $ \mathcal{H}_\tau=\mathcal{L}^2\left(-\ell_{max},+\ell_{max}\right)$.
The actions of Hermitian operators  (\ref{a9},\,\ref{a15}) in  $\mathcal{H}_\tau$ read as follows
\begin{eqnarray}\label{r1}
	\hat x\Phi(x)=x \Phi(x)\quad \mbox{and}\quad	\hat p\Phi(x)=-i\hbar D_x\Phi(x),
\end{eqnarray}
where $\Phi(x)\in \mathcal{H}_\tau$ and $D_x=\left(1-\tau  x+\tau^2 x^2\right)^{1/2}\partial_x\left(1-\tau  x+\tau^2 x^2\right)^{1/2}$ is the position-deformed derivation. Obviously, for $\tau\rightarrow 0$, we recover the ordinary  derivation. }
\end{definition}
To construct a Hilbert
space representation that describes the maximal length uncertainty  and the minimal momentum uncertainty (\ref{q2}), one has to solve the following eigenvalue problem
on the position space
\begin{eqnarray}\label{a20}
	-i\hbar D_x \Phi_\xi(x)=\xi\Phi_\xi(x),\quad\quad \xi\in\mathbb{R}.
\end{eqnarray}\label{a7}
Equation (\ref{a20}) can be conveniently rewritten by means of the transformation $ \Phi_\xi(x)=(1-\tau x+\tau^2 x^2)^{-1/2}\phi_\xi(x)$, which gives,
\begin{eqnarray}\label{a5}
	-i\hbar(1-\tau x+\tau^2 x^2)\partial_x\phi_\xi(x)=\xi\phi_\xi(x),
\end{eqnarray}
where $ \phi_\xi\in\mathcal{H}_\tau$. The  solution  of  equation (\ref{a5}) is given by 
\begin{eqnarray}
	\phi_{\xi}(x)&=& C\exp\left(i\frac{2\xi}{\tau\hbar \sqrt{3}}\left[\arctan\left(\frac{2\tau x-1}{\sqrt{3}}\right)
	+\frac{\pi}{6}\right]\right),\\
	\Phi_{\xi}(x)&=& \frac{C}{\sqrt{1-\tau x+\tau^2 x^2}}\exp\left(i\frac{2\xi}{\tau\hbar \sqrt{3}}\left[\arctan\left(\frac{2\tau x-1}{\sqrt{3}}\right)
	+\frac{\pi}{6}\right]\right),\label{a3}
\end{eqnarray}  
where $C$ is an abritrary constant. One can notice that if the standard wave-function $ \Phi_{\xi}(x)$ is normalized, then $ \phi_{\xi}(x)$ is normalized under a $\tau$-deformed integral. Indeed, we have
\begin{eqnarray}\label{Z2}
\langle \Phi_\xi|\Phi_\xi\rangle=\int_{-\ell_{max}}^{+\ell_{max}}dx \Phi^*_{\xi}(x)\Phi_{\xi}(x)=\int_{-\ell_{max}}^{+\ell_{max}}\frac{dx}{1-\tau x+\tau^2 x^2}\phi^*_\xi(x)\phi_\xi(x)=1.
\end{eqnarray}
Based on this, the normalized constant $C$ is determined as follows
\begin{eqnarray}\label{nzeta}
	C= \left(\int_{-\ell_{max}}^{+\ell_{max}}\frac{dx}{1-\tau x+\tau^2x^2}\right)^{-\frac{1}{2}}  =\sqrt{\frac{\tau\sqrt{3}}{\pi}} \label{a8}.
\end{eqnarray}
As we can see, this normalization constant \eqref{nzeta} differs from the one found in \cite{52} because of the different  boundary values $\ell_{max}$ 
 takes into account. In addition, the wavefunction  is enhanced over the one derived in  \cite{52}   by the addition of the term
  $ 1/\sqrt{1-\tau x+\tau^2 x^2}$. This correction term results from the  similarity transformation of the non-Hermitian operators to the  Hermitian operators. As a result  of this fact, Fourier transform, its inverse representation and the path integral formulation will all be improved by this correction term.  \\\\
  \begin{remark}
  	   i) From equation (\ref{Z2}), one can notice the existence of the following identity relations: 
\begin{itemize}
	\item On the $ \Phi_{\xi}(x)$-representation we have
	\begin{eqnarray}\label{id2}
		\int_{-\ell_{max}}^{+\ell_{max}}dx|x\rangle \langle x|=\mathbb{I}_{\Phi_{\xi}}.
	\end{eqnarray}
\item On the $ \phi_{\xi}(x)$-representation we have
\begin{eqnarray}
	\int_{-\ell_{max}}^{+\ell_{max}}\frac{dx}{1-\tau x+\tau^2 x^2}|x\rangle \langle x|=\mathbb{I}_{\phi_{\xi}}.
\end{eqnarray}
\end{itemize}
ii) Eigenvectors $|\Phi_\xi\rangle$ are   physically relevant i.e., there are square integrable wavefunction such that
\begin{eqnarray}\label{az1}
	||\Phi_\xi||^2=\int_{-\ell_{max}}^{+\ell_{max}} \frac{dx}{1-\tau x+\tau^2 x^2}|\phi_\xi(x)|^2<\infty.
\end{eqnarray}
iii) The  expectation values of the position energy operator  $\hat X^n$ ($n\in \mathbb{N}$)  within the states  $ |\Phi_\xi\rangle $ is finite:
\begin{eqnarray}\label{a8}
\langle \Phi_\xi|\hat X^n|\Phi_\xi\rangle = \int_{-\ell_{max}}^{+\ell_{max}} \frac{x^n dx}{1-\tau x+\tau^2 x^2}|\phi_\xi(x)|^2<\infty.
\end{eqnarray}\\
iv) The non-orthogonality relation:
\begin{eqnarray}
	\langle \Phi_{\xi'}|\Phi_{\xi}\rangle&=& \frac{\tau\sqrt{3}}{\pi}\int_{-\ell_{max}}^{+\ell_{max}}\frac{dx}{1-\tau x+\tau^2 x^2}\exp\left(i\frac{2(\xi-\xi')}{\tau\hbar \sqrt{3}}\left[\arctan\left(\frac{2\tau x-1}{\sqrt{3}}\right)+\frac{\pi}{6}\right]\right)\cr
	&=& \frac{\tau\hbar\sqrt{3}}{\pi(\xi-\xi')}\sin\left(\pi\frac{\xi-\xi'}{\tau\hbar \sqrt{3}}\right).\label{za}
\end{eqnarray}
This relation shows that, the normalized eigenstates \eqref{za} are no longer orthogonal. However, if one tends $(\xi - \xi')\rightarrow \infty $, these states become orthogonal
\begin{eqnarray}\label{a10}
\lim_{(\xi - \xi')\rightarrow \infty} \langle \Phi_{\xi'}|\Phi_{\xi}\rangle=0.
\end{eqnarray} 
For  $(\xi - \xi')\rightarrow 0,$ we have 
\begin{eqnarray}\label{aa11}
	\lim_{(\xi - \xi')\rightarrow 0} \langle \Phi_{\xi'}|\Phi_{\xi}\rangle=1.
\end{eqnarray}
These properties show that, the states $|\Phi_{\xi}\rangle$ are essentially Gaussians centered at
$ (\xi - \xi')\rightarrow 0 $ (see Figure 1). This indicates quantum fluctuations at this scale and these
fluctuations increase with the deformed parameter $\tau$.
 \begin{figure}[htbp]
	\resizebox{0.95\textwidth}{!}{
		\includegraphics {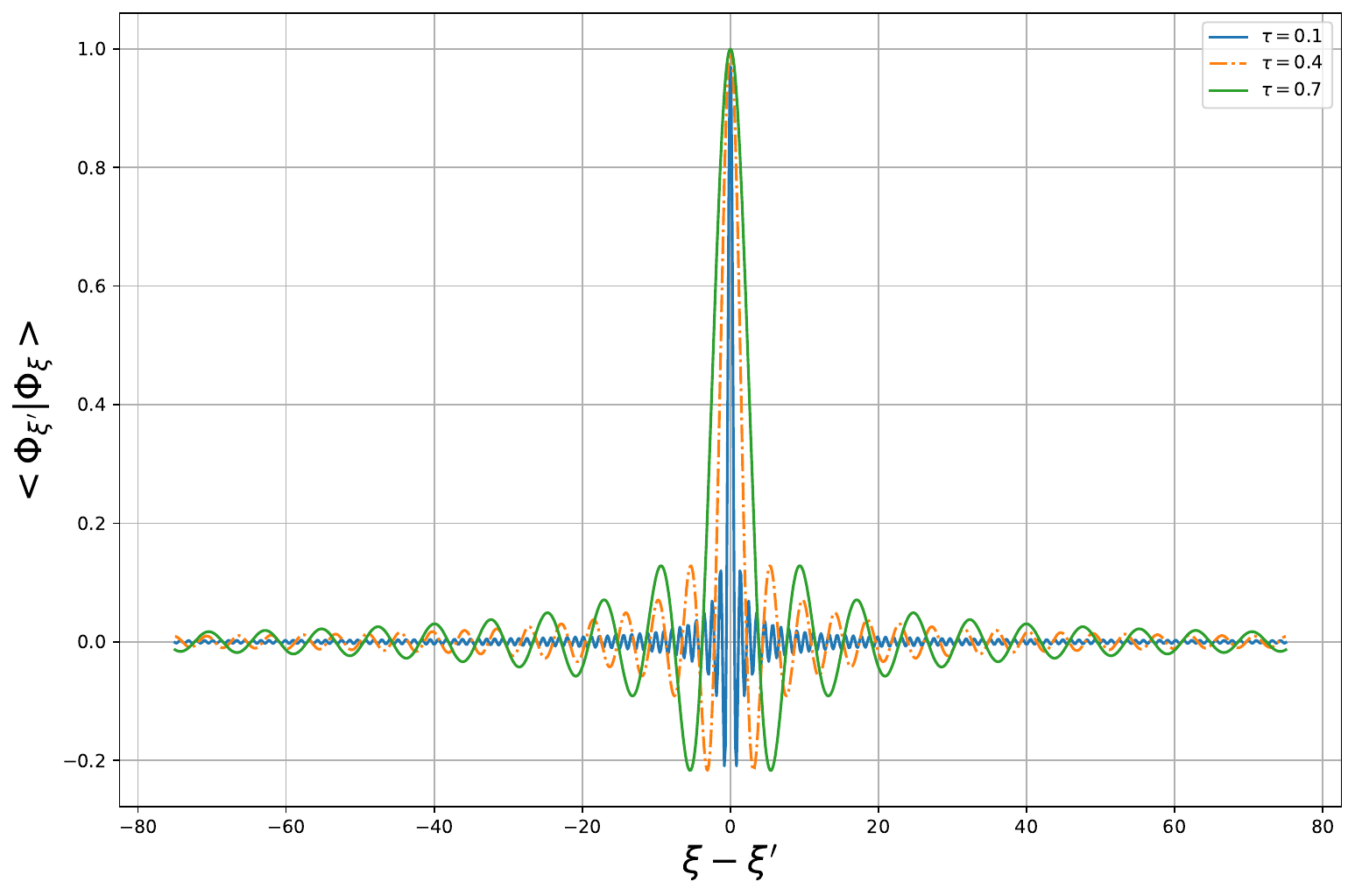}
	}
	\caption{\it \small 
	Variation of  $ \langle \Phi_{\xi'}|\Phi_{\xi}\rangle	$ versus  $\xi-\xi' $ 
	}
	\label{fig4}       
\end{figure}\\\\
v)\,The discreteness of the space:\\
Since  the scalar product \eqref{az1} vanishes in the limit $(\xi-\xi')\rightarrow \infty$, the states become orthogonal. The quantization follows from the condition 
\begin{eqnarray}\label{reslt11}
	\pi\frac{\xi-\xi'}{\tau\hbar \sqrt{3}}&=&n\pi\cr \xi-\xi'&=&\xi_n=\tau \hbar \sqrt{3}n, \quad n\in \mathbb{Z}.
\end{eqnarray}
One notices that the spectrum of momentum operator $\hat p$ presents 
discrete values. From the latter equation, one sees that 
\begin{eqnarray} \label{reslt1}
	\delta \xi_n=\xi_{n+1}-\xi_n= \tau \hbar \sqrt{3}=\sqrt{3}\Delta p_{min}.
\end{eqnarray}
With the  above results \eqref{reslt11} and \eqref{reslt1} at hand, one confirms that the formal momentum eigenvectors  $ \big|\Phi_{\xi_n}\big\rangle =\big|\Phi_{\tau \hbar \sqrt{3}n}\big\rangle $  are physically accepted and relevant. One  may be  tempted to interpret  the result (\ref{reslt1}) as if we are  describing physics on a
lattice in which   each  sites are spacing by the value $\sqrt{3}\Delta p_{min}$  illustrated as follows
	\begin{eqnarray*}
	\begin{tikzpicture}[node distance=3cm and 4cm]
		\node[] (O) at (-2,0) {$ $};
		\node[] (A) at (0,0) {$ \bullet$};
		\node (B) at (2,0) {$\bullet$};
		\node (C) at (4,0) {$\bullet$};	
		\node (D) at (6,0) {$\bullet$};
		\node (E) at (8,0) {$\bullet$};	
		\node (F) at (10,0) {$ $};
		\path[dashed,-] (O) edge (A);
		\path[-]  (A) node[below]{$\xi_1$}(A)  edge  node[above] {$\sqrt{3}\Delta p_{min}$} (B);
		\path[-] (B) node[below]{$\xi_2$} (B)  edge  node[above] {$\sqrt{3}\Delta p_{min}$}  (C);
			\path[-] (C) node[below]{$\xi_3$} (C) edge  node[above] {$\sqrt{3}\Delta p_{min}$}  (D);
			\path[-] (D) node[below]{$\xi_4$}(D) edge  node[above] {$\sqrt{3}\Delta p_{min}$}  (E);
		\path[dashed,-] (E)node[below]{$\xi_5$}(E) edge   (F);	
	\end{tikzpicture}	      
\end{eqnarray*}
 We interpret
this as the space essentially having a discrete nature. Note that similar quantization of length was shown in the context of loop quantum gravity in\cite{63,64,65,66}, albeit
following a much more involved analysis, and perhaps
under a stronger set of starting assumptions.
The wavefunctions \eqref{a3}  are square integrable functions \eqref{az1}, stable for the mean value of energy operator \eqref{a8}, have Gaussian distributions \eqref{a10} \eqref{aa11}  and  have  a discreteness nature \eqref{reslt1}.  Consequently,  the   wavefunctions \eqref{a3}  are physically  accepted  and meaningful. Its representation in  the Hermitian position-deformed Heisenberg algebra \eqref{alg5} are summarized by the following proposition.
\end{remark}
\begin{proposition}
	{\it \,\,Given a  Hilbert space $\mathcal{H}_\tau$ with the inner product $\langle .|.\rangle$, the representation of  Hermitian operators $\{\hat x,\hat p,\hat h\}$ in this space reads as  follows
\begin{eqnarray}
	\hat x\Phi_\xi(x)&=&x\Phi_\xi(x),\\
	\hat p\Phi_\xi(x)&=&-i\hbar D_x\Phi_\xi(x)=\xi \Phi_\xi(x),\\
	\hat h\Phi_\xi(x)&=&\left[-\frac{\hbar^2}{2m}D_x^2+V(x)\right]\Phi_\xi(x)= \left[-\frac{\hbar^2}{2m}\xi^2+V(x)\right]\Phi_\xi(x).\label{sc1}
\end{eqnarray}}
\end{proposition}
\begin{proof}
	The proof follows from the equations \eqref{r1} and \eqref{a20}.
\end{proof}

\subsection{ Fourier transform and its inverse
	representations }
Since the states $|\Phi_\xi\rangle$ are  physically meaningful and are well localized,  one can determine its Fourier transform (FT) and its inverse   representations  by projecting  an arbitrary state $|\Psi\rangle$.\\
\begin{definition}
 {\it Let $ \mathcal{S}\left(\mathbb{R}\right)$ be the Schwarz space which is dense in $\mathcal{H}=\mathcal{L}^2\left(\mathbb{R}\right)$. Let $|\Psi\rangle \in\mathcal{S}\left(\mathbb{R}\right)$, the FT denoted by $ \mathcal{F}_\tau[\Psi]$ or $\Psi (\xi)$  is given by
	\begin{eqnarray}
		\Psi (\xi)=\mathcal{F}_\tau[\Psi(x)](\xi)
		=\sqrt{\frac{\tau\sqrt{3}}{\pi}}  \int_{-\ell_{max}}^{+\ell_{max}}\frac{\Psi(x)dx}{\sqrt{1-\tau  x+\tau^2 x^2}} 
		e^{-i\frac{2\xi}{\tau \hbar \sqrt{3}}\left[\arctan\left(\frac{2\tau x-1}{\sqrt{3}}\right)
			+\frac{\pi}{6}\right]}.\label{moment}
	\end{eqnarray}
	The inverse  FT is given by 
	\begin{eqnarray}
		\Psi(x)=\mathcal{F}_\tau^{-1}[\Psi(\xi)](x)=\frac{1}{\hbar\sqrt{4\pi\tau\sqrt{3}}} \int_{-\infty}^{+\infty} \frac{d\xi \Psi(\xi)}{\sqrt{1-\tau  x+\tau^2 x^2}}  e^{i\frac{2\xi}{\tau \hbar \sqrt{3}}\left[\arctan\left(\frac{2\tau x-1}{\sqrt{3}}\right)
			+\frac{\pi}{6}\right]}.\label{F1}
\end{eqnarray}}
\end{definition}
\begin{remark}
i) From the  FT and inverse definitions follows the inequalities 
\begin{eqnarray}
|\Psi(\xi)|^2&\leq&\sqrt{\frac{\tau\sqrt{3}}{\pi}}  \int_{-\ell_{max}}^{+\ell_{max}} \frac{dx }{\sqrt{1-\tau  x+\tau^2 x^2}}  |\Psi(x)|^2<\infty,\\
|\Psi(x)|^2&\leq& \frac{1}{\hbar\sqrt{4\pi\tau\sqrt{3}}} \int_{-\infty}^{+\infty} \frac{d\xi }{\sqrt{1-\tau  x+\tau^2 x^2}} |\Psi(\xi) |^2<\infty.
\end{eqnarray}
ii)  As we have  mentioned, the correction factor  $ 1/\sqrt{1-\tau x+\tau^2 x^2}$ enhances  this FT and its inverse representations over the one previously obtained in \cite{67}. Therefore, on this FT representation, the action of quasi-Hermitian operators will also be modified.  
\end{remark}
\begin{Properties}
 Let   $|\Psi\rangle, |\Upsilon \rangle \in \mathcal{S}\left(\mathbb{R}\right),$  based on  the  definition of FT we have the following  properties
   \begin{align}
 &\hfill\text{i) }&  	 \mathcal{F}_\tau[\alpha\Psi(x)+\beta \Upsilon(x)](\xi)&=  \alpha\Psi(\xi)+  \beta\Upsilon(\xi),\quad \alpha,\beta\in\mathbb{C},\\
  &\hfill\text{ii) }& \frac{1}{2\hbar \tau \sqrt{3}} \int_{-\infty}^{+\infty}	 |\mathcal{F}_\tau[\Psi(x)](\xi)|^2 d\xi&=  \int_{-\ell_{max}}^{+\ell_{max}}	 |\Psi(x)|^2 dx,
      \end{align}
   where the relations (i) and (ii) are respectively the linearity  and the Parseval's identity of FT. One may also deduce the convolution property of FT. For technical reasons, we arbitrary skipe these aspects of the study and we hope to report elsewhere.
\end{Properties}
  \begin{proof}
i) For $ \alpha,\beta\in \mathbb{C}$, we have 
\begin{eqnarray*}
\mathcal{F}_\tau[\alpha\Psi(x)+\beta \Upsilon(x)](\xi)&=& \sqrt{\frac{\tau\sqrt{3}}{\pi}}  \int_{-\ell_{max}}^{+\ell_{max}}\frac{\alpha\Psi(x)dx}{\sqrt{1-\tau  x+\tau^2 x^2}} 
e^{-i\frac{2\xi}{\tau \hbar \sqrt{3}}\left[\arctan\left(\frac{2\tau x-1}{\sqrt{3}}\right)
	+\frac{\pi}{6}\right]}\cr&&+ \sqrt{\frac{\tau\sqrt{3}}{\pi}}  \int_{-\ell_{max}}^{+\ell_{max}}\frac{\beta\Upsilon(x)dx}{\sqrt{1-\tau  x+\tau^2 x^2}} 
e^{-i\frac{2\xi}{\tau \hbar \sqrt{3}}\left[\arctan\left(\frac{2\tau x-1}{\sqrt{3}}\right)
+\frac{\pi}{6}\right]}\cr
&=&\alpha\sqrt{\frac{\tau\sqrt{3}}{\pi}}  \int_{-\ell_{max}}^{+\ell_{max}}\frac{\Psi(x)dx}{\sqrt{1-\tau  x+\tau^2 x^2}} 
e^{-i\frac{2\xi}{\tau \hbar \sqrt{3}}\left[\arctan\left(\frac{2\tau x-1}{\sqrt{3}}\right)
	+\frac{\pi}{6}\right]}\cr&&+ \beta\sqrt{\frac{\tau\sqrt{3}}{\pi}}  \int_{-\ell_{max}}^{+\ell_{max}}\frac{\Upsilon(x)dx}{\sqrt{1-\tau  x+\tau^2 x^2}} 
e^{-i\frac{2\xi}{\tau \hbar \sqrt{3}}\left[\arctan\left(\frac{2\tau x-1}{\sqrt{3}}\right)
	+\frac{\pi}{6}\right]}\\
&=&\alpha \Psi(\xi)+\beta \Upsilon(\xi).
\end{eqnarray*}
ii) From the FT, we have 
\begin{eqnarray*}
 \int_{-\infty}^{+\infty}	 |\mathcal{F}_\tau[\Psi(x)](\xi)|^2 d\xi&=& \int_{-\infty}^{+\infty}|\Psi(\xi)|^2d\xi=\int_{-\infty}^{+\infty}\Psi(\xi)\Psi^*(\xi)d\xi\cr
 &=&\sqrt{\frac{\tau\sqrt{3}}{\pi}}\int_{-\infty}^{+\infty}d\xi\Psi(\xi)\cr&&\times\left[  \int_{-\ell_{max}}^{+\ell_{max}}\frac{\Psi^*(x)dx}{\sqrt{1-\tau  x+\tau^2 x^2}} 
 e^{i\frac{2\xi}{\tau \hbar \sqrt{3}}\left[\arctan\left(\frac{2\tau x-1}{\sqrt{3}}\right)
 	+\frac{\pi}{6}\right]}\right]\cr
 &=&\sqrt{\frac{\tau\sqrt{3}}{\pi}}\int_{-\ell_{max}}^{+\ell_{max}}\Psi^*(x)dx\cr&&\times\int_{-\infty}^{+\infty} \frac{\Psi(\xi)d\xi}{\sqrt{1-\tau  x+\tau^2 x^2}} 
 e^{i\frac{2\xi}{\tau \hbar \sqrt{3}}\left[\arctan\left(\frac{2\tau x-1}{\sqrt{3}}\right)
 	+\frac{\pi}{6}\right]}\cr
 &=&2\hbar \tau \sqrt{3}\int_{-\ell_{max}}^{+\ell_{max}}\Psi^*(x)\Psi(x)dx\cr
 &=&2\hbar \tau \sqrt{3}\int_{-\ell_{max}}^{+\ell_{max}}|\Psi(x)|^2dx.
\end{eqnarray*}
 \end{proof}

\begin{proposition}
	 {\it Since the states $\Phi_\xi(x)$ are  physically meaningful, there exist  a new  identity operator   defined on $\mathcal{S}$  
\begin{eqnarray}\label{id}
	\int_{-\infty }^{+\infty}\frac{d\xi}{2\hbar\tau\sqrt{3}} |\xi\rangle\langle \xi|=\mathbb{I}_{\mathcal{S}}.
\end{eqnarray}}
\end{proposition}
\begin{proof}
Using equations (\ref{a3}) and (\ref{F1}), we have
\begin{eqnarray*}
	\langle x|\Psi\rangle&=& \frac{1}{2\hbar\tau\sqrt{3}}\int_{-\infty}^{+\infty}d\xi  \langle x|\xi\rangle \langle \xi|\Psi\rangle\cr
	&=& \frac{1}{\hbar\sqrt{4\pi\tau\sqrt{3}}}\int_{-\infty}^{+\infty}d\xi\Phi_\xi  (x)\Psi(\xi)\cr
	&=& \frac{1}{\hbar\sqrt{4\pi\tau\sqrt{3}}} \int_{-\infty}^{+\infty}d\xi (1-\tau  x+\tau^2 x^2)^{-\frac{1}{2}}\Psi(\xi)  e^{i\frac{2\xi}{\tau \hbar \sqrt{3}}\left[\arctan\left(\frac{2\tau x-1}{\sqrt{3}}\right)
		+\frac{\pi}{6}\right]}, 
\end{eqnarray*}
which is equation (\ref{F1}). This confirms  the claim that equation (\ref{id}) is a correct expression for the identity  which will play the role of the completeness relation of the momentum eigenstates in the derivation of the path-integral. 
\end{proof}
\begin{corollary}
i) Let us consider arbitrary states $ |\Xi\rangle,|\Theta\rangle\in \mathcal{S}\left(\mathbb{R}\right)$, using the identity relation (\ref{id}), their scalar product reads as follows
\begin{eqnarray}
	\langle \Xi|\Theta\rangle&=& \frac{1}{2\hbar\tau\sqrt{3}}\int_{-\infty }^{+\infty} d\xi \Xi^*(\xi)\Theta(\xi),\\
	&=&\frac{1}{2\pi\hbar}\int_{-\infty}^{+\infty}d\xi\int_{-\ell_{max}}^{+\ell_{max}}\int_{-\ell_{max}}^{+\ell_{max}}\frac{dx'}{\sqrt{1-\tau  x'+\tau^2 x'^2}} \frac{dx'} {\sqrt{1-\tau  x+\tau^2 x^2}} \cr&&\times
	e^{i\frac{2\xi}{\tau \hbar \sqrt{3}}\left[\arctan\left(\frac{2\tau x-1}{\sqrt{3}}\right)
		-\arctan\left(\frac{2\tau x'-1}{\sqrt{3}}\right)
		\right]} \Xi(x') \Theta(x).
\end{eqnarray}
ii) The orthogonality of unit vector  $|x\rangle$ is given by  
\begin{eqnarray}
	\langle x|x'\rangle&=& \int_{-\infty }^{+\infty} \frac{d\xi}{2\hbar\tau\sqrt{3}}\langle x|\xi\rangle\langle \xi|x'\rangle= \int_{-\infty }^{+\infty} \frac{d\xi}{2\hbar\tau\sqrt{3}}	\Phi_\xi (x)	\Phi_\xi^* (x')\cr
		&=& \frac{1}{2\pi\hbar}\int_{-\infty }^{+\infty} d\xi \exp\left(i\frac{2\xi}{\tau\hbar \sqrt{3}}\left[\arctan\left(\frac{2\tau x-1}{\sqrt{3}}\right)
	- \arctan\left(\frac{2\tau x'-1}{\sqrt{3}}\right)\right]\right)\cr
	&=&\frac{\tau\sqrt{3}}{2}\delta\left(\arctan\left(\frac{2\tau x-1}{\sqrt{3}}\right)
	- \arctan\left(\frac{2\tau x'-1}{\sqrt{3}}\right)\right)\cr
	&=& (1-\tau  x +\tau^2  x^2)\delta(x-x').
\end{eqnarray}
\end{corollary}
\begin{proposition}
	 From the definition of FT and its inverse, it is straightfoward to show that:
\begin{align}
&\hfill\text{i) }&		\frac{d}{d\xi}\Psi(\xi)&= -i\frac{2}{\tau \hbar \sqrt{3}}\left[\arctan\left(\frac{2\tau x-1}{\sqrt{3}}\right)
	+\frac{\pi}{6}\right]\Psi(\xi),\label{key}\\
&\hfill\text{ii) }&	\frac{d}{dx}\Psi(x)&=\left(\tau\left(\frac{1}{2}-\tau x\right)+ \frac{i\xi}{\hbar}\right)\frac{\Psi(x)}{1-\tau x+\tau^2 x^2}.\label{qw}
\end{align}
\end{proposition}
\begin{lemma}
{\it The action of  Hermitian operators (\ref{moment}) on $\Psi(\xi)$  reads as follows
	\begin{eqnarray}
		\hat x\Psi(\xi)&=&\frac{2}{\tau}\frac{\tan\left(i\frac{\tau\hbar \sqrt{3}}{2}\partial_\xi\right)}{\sqrt{3}+\tan\left(i\frac{\tau\hbar \sqrt{3}}{2}\partial_\xi\right)}\Psi(\xi),\\
		\hat p \Psi(\xi)&=&\left( \xi-i2\hbar\tau\left(1-\frac{4\tan\left(i\frac{\tau\hbar \sqrt{3}}{2}\partial_\xi\right)}{\sqrt{3}+\tan\left(i\frac{\tau\hbar \sqrt{3}}{2}\partial_\xi\right)}\right)\right)\Psi(\xi),\label{qo}\\
		\hat h\Psi(\xi)&=&\frac{1}{2m}\left( \xi-i2\hbar\tau\left(1-\frac{4\tan\left(i\frac{\tau\hbar \sqrt{3}}{2}\partial_\xi\right)}{\sqrt{3}+\tan\left(i\frac{\tau\hbar \sqrt{3}}{2}\partial_\xi\right)}\right)\right)^2\Psi(\xi)\cr&&+V\left(\frac{2}{\tau}\frac{\tan\left(i\frac{\tau\hbar \sqrt{3}}{2}\partial_\xi\right)}{\sqrt{3}+\tan\left(i\frac{\tau\hbar \sqrt{3}}{2}\partial_\xi\right)}\right)\Psi(\xi).\label{sc2}
\end{eqnarray}}
\end{lemma}
\begin{proof}
	Equation (\ref{key}) is equivalent to 
	\begin{eqnarray*}
		i\frac{\tau\hbar \sqrt{3}}{2}\frac{d}{d\xi}= \left[\arctan\left(\frac{2\tau x-1}{\sqrt{3}}\right)
		+\frac{\pi}{6}\right]=
		\left[\arctan\left(\frac{2\tau x-1}{\sqrt{3}}\right)
		+\arctan\left(\frac{1}{\sqrt{3}}\right)\right].\label{eq}
	\end{eqnarray*} 
	From the following relation \cite{52}
	\begin{eqnarray*}
		\arctan\alpha +\arctan \beta=\arctan \left(\frac{\alpha+\beta}{1-\alpha\beta}\right),\quad \mbox{with}\quad \alpha\beta<1,
	\end{eqnarray*}
	we deduce that
	\begin{eqnarray*}
		\tan \left[\arctan\left(\frac{2\tau x-1}{\sqrt{3}}\right)
		+\arctan\left(\frac{1}{\sqrt{3}}\right)\right]=\frac{\tau x \sqrt{3} }{2-\tau x}.
	\end{eqnarray*}
	Therefore, the position operator $\hat x$ is represented as follows 
	\begin{eqnarray*}
		\hat x&=&\frac{2}{\tau}\frac{\tan\left(i\frac{\tau\hbar \sqrt{3}}{2}\partial_\xi\right)}{\sqrt{3}+\tan\left(i\frac{\tau\hbar \sqrt{3}}{2}\partial_\xi\right)}\mathbb{I},\\
		x\Psi(\xi)&=&\frac{2}{\tau}\frac{\tan\left(i\frac{\tau\hbar \sqrt{3}}{2}\partial_\xi\right)}{\sqrt{3}+\tan\left(i\frac{\tau\hbar \sqrt{3}}{2}\partial_\xi\right)}\Psi(\xi).
	\end{eqnarray*}
	Using equation (\ref{qw}), the action of $\hat p$ on the  quasi-representation (\ref{F1}) reads as follows
	\begin{eqnarray}
		\hat p\Psi(x)&=& \frac{-i\hbar}{\hbar\sqrt{4\pi\tau\sqrt{3}}} \int_{-\infty}^{+\infty}d\xi \Psi(\xi)\cr&&\times D_x\left( (1-\tau  x+\tau^2 x^2)^{-1/2} e^{i\frac{2\xi}{\tau \hbar \sqrt{3}}\left[\arctan\left(\frac{2\tau x-1}{\sqrt{3}}\right)
			+\frac{\pi}{6}\right]}\right)\cr
		&=& \frac{1}{\hbar\sqrt{4\pi\tau\sqrt{3}}} \int_{-\infty}^{+\infty} \left(-i\hbar\tau\left(\frac{1}{2}-\tau x\right)+ \xi\right) \Psi(\xi)\cr&&\times\frac{d\xi }{\sqrt{1-\tau  x+\tau^2 x^2}}  e^{i\frac{2\xi}{\tau \hbar \sqrt{3}}\left[\arctan\left(\frac{2\tau x-1}{\sqrt{3}}\right)
			+\frac{\pi}{6}\right]}.\label{q1}
	\end{eqnarray}
On the other hand, the action of $\hat p$ on the  quasi-representation (\ref{F1}) reads as follows
	\begin{eqnarray}
		\hat p\Psi(x)&=& \frac{1}{\hbar\sqrt{4\pi\tau\sqrt{3}}} \int_{-\infty}^{+\infty}\hat p \Psi(\xi) \cr&&\times\frac{d\xi }{\sqrt{1-\tau  x+\tau^2 x^2}}  e^{i\frac{2\xi}{\tau \hbar \sqrt{3}}\left[\arctan\left(\frac{2\tau x-1}{\sqrt{3}}\right)
			+\frac{\pi}{6}\right]}.\label{b5}
	\end{eqnarray}
	By comparing equation (\ref{q1}) and equation (\ref{b5}), we obtain equation (\ref{qo}) of Lemma 4.2	
	\begin{eqnarray*}
		\hat p \Psi(\xi)&=&\left( \xi-i\hbar\tau\left(\frac{1}{2}-\tau x\right)\right) \Psi(\xi)\cr
		&=&\left( \xi-i2\hbar\tau\left(1-\frac{4\tan\left(i\frac{\tau\hbar \sqrt{3}}{2}\partial_\xi\right)}{\sqrt{3}+\tan\left(i\frac{\tau\hbar \sqrt{3}}{2}\partial_\xi\right)}\right)\right)\Psi(\xi).
	\end{eqnarray*}
	The Hamiltonian is given  by
	\begin{eqnarray*}
		\hat h\Psi(\xi)&=&\left(\frac{\hat p^2}{2m}+V(\hat x)\right)\Psi(\xi)
		=\frac{1}{2m}\left( \xi-i2\hbar\tau\left(1-\frac{4\tan\left(i\frac{\tau\hbar \sqrt{3}}{2}\partial_\xi\right)}{\sqrt{3}+\tan\left(i\frac{\tau\hbar \sqrt{3}}{2}\partial_\xi\right)}\right)\right)^2\Psi(\xi)\cr&&+V\left(\frac{2}{\tau}\frac{\tan\left(i\frac{\tau\hbar \sqrt{3}}{2}\partial_\xi\right)}{\sqrt{3}+\tan\left(i\frac{\tau\hbar \sqrt{3}}{2}\partial_\xi\right)}\right)\Psi(\xi).
	\end{eqnarray*}
\end{proof}
\begin{remark}
From the limit $\tau\rightarrow 0$ in the last equations, we recover the  ordinary  representations in momentum space as
\begin{eqnarray}
	\lim_{\tau\rightarrow 0}\hat x\Psi(\xi)&=&i\hbar\partial_\xi\Psi(\xi),\\
	\lim_{\tau\rightarrow 0}	\hat p \Psi(\xi)&=&\xi\Psi(\xi),\\
	\lim_{\tau\rightarrow 0} \hat h\Psi(\xi)&=&\left(\frac{\xi^2}{2m}+V\left(i\hbar\partial_\xi\right)
	\right)\Psi(\xi).
\end{eqnarray}
\end{remark}

\section{Path integral }\label{sec5}
From the path integrals within this position-deformed Heisenberg algebra, we construct the propagator depending on the position-representation and on the Fourier transform and its inverse representations.  We compute propagators and deduce the actions of a free particle. 
\subsection{Path integral  in position-space representation}
 \begin{definition}
  {\it \, The path integral is defined by 
\begin{eqnarray}\label{F31}
	\Phi_\xi (x,t)=  \int_{-l_{max}}^{+l_{max} }dx'K(x,x',\Delta t) \Phi_\xi(x',t'),
\end{eqnarray}
where $K$ is the kernel in the Hamiltonian  or 
the amplitude for a particle to propagate from the state with position $x'$ to the state with position $x\,  (x>x')$ in a time interval  $\Delta t=t-t'  $ \cite{68,69}   and it  is  defined as
\begin{eqnarray}\label{path1}
	K(x,x', \Delta t)=\langle x|e^{-\frac{i}{\hbar}\hat h\Delta t} |x'\rangle.
\end{eqnarray}}
\end{definition}
\begin{proposition}
 {\it As easily checked the  kernel (\ref{path1}) satisfies the following equations:\\
\begin{align}
&\hfill\text{i) }&	-\frac{\hbar^2}{2m} D_{x'}^2 K(x,x', \Delta t) +V(x')K(x,x', \Delta t)= i\hbar \partial_t K(x,x', \Delta t),\\	
&\hfill\text{ii) }& K(x,x', 0)=(1-\tau  x +\tau^2  x^2)\delta(x-x'),\\
&\hfill\text{iii) }& \int_{-\ell_{max}}^{+\ell_{max} }dx''	K(x,x'',\Delta t_1 ) K(x'',x',\Delta t_2 )= K(x,x',\Delta t_1+\Delta t_2 ),\\
&\hfill\text{iv)}& K^\dag(x,x', \Delta t)= K(x',x, -\Delta t),
\end{align}}
where these equations are respectively: i) Schrödinger equation; ii) Initial  condition; iii) Composition rule; iv) Unitarity. 
\end{proposition}
\begin{proof}
i) $i\hbar \partial_t K(x,x', \Delta t)=\langle x|i\hbar \partial_te^{-\frac{i}{\hbar}\hat h\Delta t} |x'\rangle=\langle x|\hat h e^{-\frac{i}{\hbar}\hat h\Delta t} |x'\rangle=\langle x| e^{-\frac{i}{\hbar}\hat h\Delta t}\hat h |x'\rangle=\\h( p, x' ) \langle x|e^{-\frac{i}{\hbar}\hat h\Delta t} |x'\rangle= h( p, x' )K(x,x', \Delta t) =\left(	-\frac{\hbar^2}{2m} D_{x'}^2  +V(x')\right) K(x,x', \Delta t)$.    \\
ii) $K(x,x', 0)= \langle x|x'\rangle$. Referring to the equation \eqref{qw}, we have $K(x,x', 0)= \langle x|x'\rangle=(1-\tau  x +\tau^2  x^2)\delta(x-x')$. iii) $K(x,x',\Delta t_1+\Delta t_2 )=\langle x|e^{-\frac{i}{\hbar}\hat h(\Delta t_1+\Delta t_2)} |x'\rangle$\\= $\langle x|e^{-\frac{i}{\hbar}\hat h\Delta t_1 } e^{-\frac{i}{\hbar}\hat h\Delta t_2 } |x'\rangle=\int_{-\ell_{max}}^{+\ell_{max} }dx'' \langle x|e^{-\frac{i}{\hbar}\hat h\Delta t_1 }|x''\rangle \langle x''| e^{-\frac{i}{\hbar}\hat h\Delta t_2 } |x'\rangle$\\$= \int_{-\ell_{max}}^{+\ell_{max} }dx'' K(x,x'', \Delta t_1)K(x'',x', \Delta t_2)$.  iv) $ K^\dag(x,x', \Delta t)=\langle x'|e^{\frac{i}{\hbar}\hat h\Delta t} |x\rangle= K^\dag(x',x, -\Delta t).  $
\end{proof}

Splitting the interval $t-t'$ into $N$ intervals of length $\epsilon=(t_k-t_{k-1})/N$ and inserting the completeness relations in (\ref{id2}) and (\ref{id}), the propagator (\ref{path1}) becomes
\begin{eqnarray}\label{prope}
	K(x,x',\Delta t)&=&  \int_{-l_{max}}^{+l_{max}} \left(\prod_{k=1}^{N-1} dx_k\right) \int_{-\infty}^{+\infty} \left( \prod_{k=1}^{N} \frac{d\xi_k}{2\pi\hbar\tau\sqrt{3}} \right)\cr&&\times \langle x_k|\xi_k\rangle\langle \xi_k|e^{-\frac{i}{\hbar}\epsilon\hat h}|x_{k-1}\rangle.
\end{eqnarray}
Recall that
\begin{eqnarray}
	\langle x_k|\xi_k\rangle&=& \Phi_{\xi_k}(x_k) = \frac{\sqrt{\frac{\tau\sqrt{3}}{\pi}}}{\sqrt{1-\tau x_k+\tau^2 x_k^2}} e^{\left(i\frac{2\xi_k}{\tau\hbar \sqrt{3}}\left[\arctan\left(\frac{2\tau x_k-1}{\sqrt{3}}\right)
		+\frac{\pi}{6}\right]\right)}, \label{z4}\\
	\langle \xi_k|e^{-\frac{i}{\hbar}\epsilon\hat h}|x_{k-1}\rangle & \simeq& e^{-\frac{i}{\hbar}\epsilon h(\xi_k,x_{k-1})}  \langle \xi_k|x_{k-1}\rangle+\mathcal{O}(\epsilon^2) \cr
	&\simeq& e^{-\frac{i}{\hbar}\epsilon h(\xi_k,x_{k-1})} \Phi_{\xi_k}^*(x_{k-1})+ \mathcal{O}(\epsilon^2).\label{z5}
\end{eqnarray}
\begin{proposition}
{\it \,
Substituting equations (\ref{z4}) and (\ref{z5}) into  equation (\ref{prope}) gives the discrete propagator
\begin{eqnarray}\label{q21}
	K_{disc}(x,x',\Delta t)&=&\left[ \int_{-l_{max}}^{+l_{max}} \left(\prod_{k=1}^{N-1}\frac{dx_k}{\sqrt{1-\tau x_k+\tau^2 x_k^2}\sqrt{1-\tau x_{k-1}+\tau^2 x_{k-1}^2}}\right)\right]\cr&&\times\left[  \int_{-\infty}^{+\infty} \left( \prod_{k=1}^{N}  \frac{d\xi_k}{2\pi\hbar} \right)  \right]
	e^{\frac{i}{\hbar}\epsilon\mathcal{S}_{disc}},
\end{eqnarray}
where the  discrete action $S_{disc}$ is given  by
\begin{eqnarray}
	S_{disc}= \sum_{k=1}^{N-1}\frac{2\xi_k}{\tau\sqrt{3}}\left[\frac{\arctan \left(\frac{2\tau x_{k}-1}{\sqrt{3}}\right) 
		-\arctan \left(\frac{2\tau x_{k-1}-1}{\sqrt{3}}\right)}{\epsilon}\right] -\sum_{k=1}^{N-1}h(\xi_k,x_{k-1}).
\end{eqnarray}}
\end{proposition}

\begin{lemma}
{\it \, Taking  $N\rightarrow \infty$ in equation \eqref{q21},  so that $\epsilon \rightarrow 0$ we obtain the continuous propagator as  follows
\begin{eqnarray}\label{propa}
K(x,x',\Delta t)= \int\mathcal{D}x\mathcal{D}\xi e^{\frac{i}{\hbar} S },
\end{eqnarray}
where the  integration measures $\mathcal{D}x$ and $\mathcal{D}\xi$ are defined as
\begin{eqnarray}\label{int1}
	\mathcal{D}x=\lim_{N\rightarrow \infty}\prod_{k=1}^{N-1}\frac{dx_k}{\sqrt{1-\tau x_k+\tau^2 x_k^2}\sqrt{1-\tau x_{k-1}+\tau^2 x_{k-1}^2}}\quad \mbox{and}\quad \mathcal{D}\xi=\lim_{N\rightarrow \infty} \prod_{k=1}^{N}
	\left(\frac{d\xi_k}{2\pi\hbar}\right).
\end{eqnarray}
and the continuous  action $S$ is given by
\begin{eqnarray}\label{path}
	S \left[x(t),x(t')\right]&=&\int_{t'}^t d\nu  \left[\frac{\dot{x}(\nu)}{1-\tau x(\nu)+\tau^2 x^2(\nu)}\xi(\nu)-h(\xi(\nu),x(\nu))\right],
\end{eqnarray}
where $\dot{x}(\nu)= d x/d\nu  $.}
\end{lemma}
\begin{remark} 
i) As we can see, this formulation of  path integral   is  similar to that in reference \cite{52}. This similarity arises from the realization of this formulation within the   Hermitian Heisenberg algebra \eqref{alg5}, which is equivalent to the one  used in \cite{52}. Clairy, the  Hermitian Hamiltonian variable $h(x,\xi)$, which generalizes the pseudo-Hermitian one $H(x,\xi=\rho)$ used in \cite{52}, is also present in this path integral.   
 Furthermore, as we can notice the propagator \eqref{propa} and the action \eqref{path} result from no approximation methods and allow an effective description compared to those of the references\cite{54,55,56,57}.  
\\
iii) Taking the limit $\tau\rightarrow 0$ in equation \eqref{q2}, the deformed propagator (\ref{propa}) is reduced to the ordinary one of Euclidean space such that
\begin{eqnarray}
	K^0(x,x',\Delta t)= \int\lim_{N\rightarrow \infty}\prod_{k=1}^{N-1}dx_k\prod_{k=1}^{N}
	\left(\frac{d\xi_k}{2\pi\hbar}\right)e^{\frac{i}{\hbar}S^0},
\end{eqnarray}
where  the undeformed action $S^0$ is given by 
\begin{eqnarray}\label{a40}
	S^0\left[x(t),x(t')\right]=\int_{t'}^t d\nu  \left[\dot{x}(\nu)\xi(\nu)-h(\xi(\nu),x(\nu))\right].
\end{eqnarray}
\end{remark}
\begin{theorem}{\it \,
It is straightforward to show that  the following relations
\begin{eqnarray}\label{res1}
	K(x,x',\Delta t)\leq K^0(x,x',\Delta t)\implies S\leq S^0. 
\end{eqnarray}}
\end{theorem}
\begin{proof}
	The proof follows from a  straightforward comparaison between  equations \eqref{r1} and \eqref{a20} on one hand and,  equations \eqref{path} and   \eqref{a40} on the other.
\end{proof}
It is  well known that, the action in classical mechanics is  a functional over paths that describe what is the motion of a system over a particular path. As we can see from this result \eqref{res1}, the  deformed action $S$ is  bounded by the ordinary one $S^0$ of classical mechanics. It makes sense to think of deformation effects as shortening the classical system's path, which enables quick motion in this space.

The stationary path (\ref{path}) is obtained by using the variational principle
\begin{eqnarray}\label{variational}
	\delta  S
	=\delta \int_{t'}^t d\nu L \left[\dot{x}(\nu),x(\nu)\right]=\int_{t'}^t d\nu \left(\frac{\partial L}{\partial x(\nu)}\delta x(\nu) +  \frac{\partial L}{\partial \dot{x}(\nu)}\delta \dot{x}(\nu)\right)=0,
\end{eqnarray}
where the Lagrangian $L$  of the system is given by
\begin{eqnarray}
	L \left[  \dot{x}(\nu),x(\nu)\right]= \frac{\dot{x}(\nu)}{1-\tau x(\nu)+\tau^2 x^2(\nu)}\xi(\nu)-h(\xi(\nu),x(\nu)).
\end{eqnarray}
The solutions of  equation (\ref{variational}) generate the  following differential equations
\begin{eqnarray}
	\dot{x}&=&(1-\tau x+\tau^2 x^2)\frac{\partial h}{\partial \xi}=\{x,\xi\}_\tau \frac{\partial h}{\partial \xi},\\
	\dot{\xi}&=&-(1-\tau x+\tau^2 x^2)\frac{\partial h}{\partial x}=-\{x,\xi\}_\tau \frac{\partial h}{\partial x},
\end{eqnarray}
where $\{x,\xi\}_\tau= (1-\tau x+\tau^2 x^2)$ is the position-deformed Poisson bracket.
By taking the limit $\tau\rightarrow 0$, we recover the ordinary  Hamilton equations of motion.

\subsection{Path integral in Fourier  transform and its inverse representions }
Using the generalized Fourier transform and its inverse representations (\ref{moment}), (\ref{id}) and taking into account equation (\ref{F31}), we have
\begin{eqnarray}
	\Psi (\xi,t)
	&=&
	\sqrt{\frac{\tau\sqrt{3}}{\pi}}  \int_{-\ell_{max}}^{+\ell_{max}}\frac{\Psi(x)dx}{\sqrt{1-\tau  x+\tau^2 x^2}} 
	e^{-i\frac{2\xi}{\tau \hbar \sqrt{3}}\left[\arctan\left(\frac{2\tau x-1}{\sqrt{3}}\right)
		+\frac{\pi}{6}\right]}\cr&&
	\times \int_{-l_{max}}^{+l_{max} }\frac{K(x,x',\Delta t)}{\sqrt{1-\tau  x'+\tau^2 x'^2}}dx' \frac{1}{\hbar\sqrt{4\pi\tau\sqrt{3}}} \cr&&
	\times \int_{-\infty}^{+\infty}d\xi'  e^{i\frac{2\xi'}{\tau \hbar \sqrt{3}}\left[\arctan\left(\frac{2\tau x'-1}{\sqrt{3}}\right)
		+\frac{\pi}{6}\right]}\Psi(\xi',t').
\end{eqnarray}
This path integral can be  rewritten as follows
\begin{eqnarray}
	\Psi (\xi,t)=\int_{-\infty}^{+\infty}d\xi'\mathcal{K}(\xi,\xi',\Delta t) 	\Psi (\xi',t'),
\end{eqnarray}
where $ \mathcal{K}$ is the propagator in Fourier  transform and its inverse representions for a particle to go from a state $\Psi (\xi') $
to a state $\Psi (\xi)$ in a time
interval $\Delta t$ is
\begin{eqnarray}\label{key1}
	\mathcal{K}(\xi,\xi',\Delta t)&=&\frac{1}{2\pi\hbar} \int_{-l_{max}}^{+l_{max} }\frac{dx}{\sqrt{1-\tau x+\tau^2 x^2}}\frac{dx'}{\sqrt{1-\tau x'+\tau^2 x'^2}}\cr&&\times e^{-i\frac{2}{\tau \hbar \sqrt{3}}\left[\xi\arctan\left(\frac{2\tau x-1}{\sqrt{3}}\right)
		-\xi'\arctan\left(\frac{2\tau x'-1}{\sqrt{3}}\right)\right]}
	K(x,x',\Delta t),\cr
	&=&\frac{1}{2\pi\hbar}\int\mathcal{D}x\mathcal{D}\xi\frac{dx}{\sqrt{1-\tau x+\tau^2 x^2}}\frac{dx'}{\sqrt{1-\tau x'+\tau^2 x'^2}}e^{\frac{i}{\hbar}\mathcal{S}},
\end{eqnarray}
with the  functional action $\mathcal{S}$   given by
\begin{eqnarray}
	\mathcal{S}(\xi,\xi')= S-\frac{2}{\tau  \sqrt{3}}\left[\xi\arctan\left(\frac{2\tau x-1}{\sqrt{3}}\right)
	-\xi'\arctan\left(\frac{2\tau x'-1}{\sqrt{3}}\right)\right].	
\end{eqnarray}

\subsection{Propagators for a free particle  }
The Hamiltonian of a free particle  is given by
\begin{eqnarray}
	\hat h_{fp}=\frac{\hat p^2}{2m}.
\end{eqnarray}
The propagator in position-represention in the  time interval $\Delta t =t-t'$ is  given by
\begin{eqnarray}
	K_{fp}(x,x',\Delta t) &=&\langle x|e^{-\frac{i}{\hbar}\frac{\hat p^2}{2m}\Delta t}|x'\rangle\cr
	&=&\frac{1}{2\pi\hbar\tau\sqrt{3}} \int_{-\infty}^{+\infty}d\xi e^{-\frac{i}{\hbar}\frac{\xi^2}{2m}\Delta t}\Phi_\xi(x)\Phi_\xi^*(x)\cr
	 &=&\int_{-\infty}^{+\infty}\frac{d\xi}{2\pi\hbar} e^{\left(i\frac{2\xi}{\tau\hbar \sqrt{3}}\left[\arctan\left(\frac{2\tau x-1}{\sqrt{3}}\right)
		-\arctan\left(\frac{2\tau x'-1}{\sqrt{3}}\right)\right]-\frac{i}{\hbar}\frac{\xi^2}{2m}\Delta t\right)}.\label{a11}
\end{eqnarray}
\begin{lemma}{\it\,
 Completing this Gaussian integral (\ref{a11}), the deformed-propagator $ K_{fp} $, the deformed-action $ S_{fp} $ and the deformed-kinetic energy $ T $ read as follows
\begin{eqnarray}
	K_{fp}(x,x',\Delta t) &=&	\sqrt{\frac{ m}{2\pi\hbar i\Delta t}} 	e^{i\frac{2m}{\hbar3\tau^2 \Delta t}\left[\arctan\left(\frac{2\tau x-1}{\sqrt{3}}\right)
		-\arctan\left(\frac{2\tau x'-1}{\sqrt{3}}\right)\right]^2},\label{A23}\\
	S_{fp} &=&\frac{2m}{3\tau^2 \Delta t}\left[\arctan\left(\frac{2\tau x-1}{\sqrt{3}}\right)
	-\arctan\left(\frac{2\tau x'-1}{\sqrt{3}}\right)\right]^2,\label{A22}\\
	T &=&\frac{2m}{3\tau^2 (\Delta t)^2}\left[\arctan\left(\frac{2\tau x-1}{\sqrt{3}}\right)
	-\arctan\left(\frac{2\tau x'-1}{\sqrt{3}}\right)\right]^2.\label{A24}
\end{eqnarray}}
\end{lemma}
\begin{proof}
See \cite{52} for the proof of this Lemma 5.3. 
\end{proof}
Taking the limit $\tau\rightarrow 0$ in equations \eqref{A22}, \eqref{A23} and \eqref{A24}, these equations  properly
reduce to the well-known  result in ordinary quantum mechanics for a free particle \cite{68,69}   that is
\begin{eqnarray}
	\lim_{\tau\rightarrow 0}	K_{fp}(x,x',\Delta t)&=&K_{fp}^0(x,x',\Delta t)=\sqrt{\frac{ m}{2\pi\hbar i\Delta t}} e^{\frac{i}{\hbar}\frac{m(x-x')^2}{2\Delta t}},\label{z1}\\
	\lim_{\tau\rightarrow 0} S_{fp}&=& S_{fp}^0= \frac{m}{2}\frac{(x-x')^2}{\Delta t},\label{z2}\\
	\lim_{\tau\rightarrow 0} T&=&T^0=\frac{m}{2}\frac{(x-x')^2}{(\Delta t)^2}.\label{z3}
\end{eqnarray}
\begin{theorem} {\it
It is straightforward to show  the following relations
\begin{eqnarray}\label{res}
	K_{fp}(x,x',\Delta t)\leq K_{fp}^0(x,x',\Delta t)\implies S_{fp}\leq S_{fp}^0 \implies  T\leq T^0.
\end{eqnarray}}
\end{theorem}
\begin{proof}
	The proof follows from a  straightforward comparaison between  equations of Lemma 5.3 on one hand  and equations  \eqref{z1}, \eqref{z2} and   \eqref{z3} on the other hand.
\end{proof}

This    indicates that the deformed propagator and    actions of the free particle are dominated by the standard ones without quantum  deformation.
These results indicate that the quantum deformation effects in this space  shortens the paths of particles, allowing them to move from one point to another in a short time.  In one way or another, as one can see from  equation  (\ref{res}), these results can be understood as free particles use low kinetic  energies to travel faster in this  deformed space.
This confirms our recent results \cite{49,50}  and strengthens the claim that the position deformed-algebra (\ref{alg5})  induces strong deformation of the quantum  levels allowing particles to jump from one state to another with low energy transitions \cite{49,50}.

\begin{lemma}{\it \,\, The propagator in the FT representation  is given by
\begin{eqnarray}
	\mathcal{K}_{fp}(\xi,\xi',\Delta t)&=& 
	\frac{1}{2\pi\hbar}\sqrt{\frac{ m}{2\hbar\pi i\Delta t}} \int_{-l_{max}}^{+l_{max}}\int_{-l_{max}}^{+l_{max} }\frac{dx}{\sqrt{1-\tau x+\tau^2 x^2}}  \frac{dx'}{\sqrt{1-\tau x'+\tau^2 x'^2}}\cr&&\times
	e^{\frac{i}{\hbar}\mathcal{S}_{fp}},
\end{eqnarray}
where $\mathcal{S}_{fp}$ is the corresponding action  given by
\begin{eqnarray}
	\mathcal{S}_{fp}= S_{fp}- \frac{2}{\tau  \sqrt{3}}\left[\xi\arctan\left(\frac{2\tau x-1}{\sqrt{3}}\right)
	-\xi'\arctan\left(\frac{2\tau x'-1}{\sqrt{3}}\right)\right].
\end{eqnarray}}
\end{lemma}
\begin{proof}
	See \cite{52} for the proof of this Lemma 5.4. 
\end{proof}
\section{Conclusion}\label{sec7}
The Hamiltonian operator in the study of dynamical quantum systems needs to be Hermitian. Therefore, the orthoganility of the Hamiltonian eigenbasis, the conservation of probability density, and the realism of the spectrum are all guaranteed by the Hamiltonian's Hermicity.  Within a position-deformed Heisenberg algebra \eqref{alg4}, we have demonstrated in the current study that a Hamiltonian operator with real spectrum is no longer Hermitian.  Using a similarity transformation and a suitable positive-definite Dyson map \eqref{dyson} derived from a metric operator \eqref{b1}, we have determined the Hermicity of this operator. Next, we constructed Hilbert space representations  associated with these Hermitian operators that form a  Hermitian position deformed Heisenberg algebra \eqref{alg5}. With the help of these representations we establish  path integral formulations of any systems in this  Hermitian algebra. The propagator is then considered as an example together with the appropriate action of a free particle. As a result of the Euclidean space's deformation, we have demonstrated that the action that characterizes the system's classical trajectory is constrained by the standard one of classical mechanics. Consequently,
particles of this system travel quickly from one point to another with low kinetic  energy.   Moreover, this path integral formulation generalizes the  standard formulation \cite{53} and offers an additional method to find the propagator and the action of deformed quantum  systems. In addition, this formulation requires no approximation techniques to compute the propagator and the  action of quadratic deformed systems, in contrast to formulations found in the literature \cite{54,55,56,57}. 	
	 This makes it to be useful method because it allows an effective description    dynamical   systems. This makes it a valuable approach since it enables a dynamic system to be described effectively.

Overall, the result achieved in this study is now identical to the result that was recently derived \cite{52}. This  result improves the previous one by the use of similarity transformation that restores the Hermicity of the Hamiltonian operator. It is possible to interpret the expansion of the expression $ (1-\tau x+\tau^2 x^2)^{-1/2}$ above the one obtained in \cite{52} as an improvement of  the wavefunction \eqref{a3}, the Fourier transform \eqref{moment}, and its inverse representations \eqref{F1}.
 The equivalence between the position-deformed Heisenberg algebra \cite{52} and the Hermitian position deformed Heisenberg algebra \eqref{alg5} accounts for the similarity of path formulations of a free particle for both outcomes. In summary, the current paper's finding, which was reached through the application of similarity  transformation, provides an additional method for  restoring  the loss Hermicity   and obtaining the previous on \cite{52}.

\section*{Acknowledgments}
 We would like to thank the referees for giving such constructive comments which considerably
	improved the quality of the paper. LML is fully supported by funding through AIMS-Ghana and  AIMS Research and Innovation Center (RIC).   He would like to extend his heartfelt gratitude  to Max Planck Institute  for the Science of Light for providing him with the opportunity to conduct a part of this work as a visiting researcher in Flore Kunst's research group through DAAD  visiting research grant.
	 He is  deeply grateful for the support and resources made available to him during his stay  wi, which have been instrumental in the progress of this work. 

%

%

\end{document}